\documentclass[aps,pra,twocolumn,superscriptaddress,showpacs,amsmath,amssymb]{revtex4}

\usepackage{graphicx}

\usepackage{amsfonts}

\begin{document}

\title{ Relaxation due to random collisions with a many-qudit environment}
\author{Giuseppe Gennaro}
\affiliation{  Dipartimento  di Scienze Fisiche ed Astronomiche, Universit\`a di Palermo, via Archirafi 36,
I-90123 Palermo, Italy}

\author{Giuliano Benenti}
\affiliation{CNISM, CNR-INFM \& Center for Nonlinear and Complex systems,
Universit\`{a} degli Studi dell'Insubria, via Valleggio 11,
I-22100 Como, Italy \\ \& Istituto Nazionale di Fisica Nucleare, Sezione di Milano, via Celoria 16, I-20133 Milano, Italy}

\author{G. Massimo Palma}
\affiliation{ NEST - CNR (INFM) \& Dipartimento  di Scienze Fisiche ed Astronomiche, Universit\`a di Palermo, via Archirafi 36,
I-90123 Palermo, Italy}

\begin{abstract}
We analyze the dynamics of a system qudit of dimension $\mu$
sequentially interacting 
with the $\nu$-dimensional qudits of 
a chain playing the role of an environment.  Each pairwise collision 
has been modeled as a random unitary transformation. 
The relaxation to equilibrium of the purity of the system qudit,
averaged over random collisions, is analytically computed by means 
of a Markov chain approach.
In particular, we show that the steady state is the one corresponding 
to the steady state for random collisions with a single environment
qudit of effective dimension $\nu_e=\nu\mu$.
Finally, we numerically investigate aspects 
of the entanglement dynamics for qubits 
($\mu=\nu=2$) and show
that random unitary collisions can 
create multipartite entanglement between 
the system qudit and the qudits of the chain.
\end{abstract}

\date{\today}

\maketitle

\section{Introduction}

The repeated collision model is a simple yet instructive model of irreversible quantum dynamics. First introduced in~\cite{Scarani2002} to analyze the process of thermalization and more generally of homogenization~\cite{Ziman2002,Ziman,Koniorczyk} in the limit of an environment with a large number of degrees of freedom, it  has been further studied to elucidate various aspects of the irreversible dynamics of quantum systems in the presence of environments with few degrees of freedom~\cite{Benenti2007, Gennaro2008}. In most of the literature cited above the interaction between system and environment is due to pairwise elastic collisions modeled by partial swap operators. In the present paper we will instead model such collisions by random unitary operators~\cite{Diaconis2005, Mezzadri2007, Kus1991, Kus1994, Kus1996, Kus1998}. Such choice is due to several reasons. On the one hand, although an exact modelization of the system-environment interaction is difficult, a good description of the relaxation process can be obtained by a suitable average over random interactions. Some examples of such approach can be found in~\cite{Pineda2007,Petruccione2007}, where the irreversible dynamics of a single and a pair of qubits is analyzed in terms of a random interaction  with an environment which is itself modeled as a random matrix and in~\cite{Gennaro2008}, where the average dynamics of a single qubit interacting with a very small reservoir is described again by random collisions. On the other hand random and pseudo-random states and their efficient generation by suitable sequences of random gates~\cite{Emerson2003, Emerson2004, Emerson2005, Weinstein2005,
Plenio2007,Oliveira2007,Znidaric2007,Rossini2008,Benenti2008,Znidaric2008} have received a considerable attention due to possible applications in quantum information processing~\cite{Harrow2004,Bennett2004,Hayden2004,Hayden2006}.

In the present paper we analyze the approach to equilibrium of a system qudit interacting with a very large ensemble of qudits. The interaction is modeled  by a sequence of  two-qudit random collisions. The paper is structured as follows: in the following section we review the random collision model and we specialize it to the case of collisions described by random unitary operators. We then characterize the approach to equilibrium by analytically calculating  the purity of the system steady state and the rate of approach to such state for the specific case of colliding qubits. Such analytical analysis is then generalized to a system qudit of dimension $\mu$ colliding with an ensemble of qudits of size $\nu$. We then proceed with a numerical analysis of the entanglement dynamics for colliding qubits

\section{The random collision model and the relaxation to equilibrium}
In the random collisions model of irreversible dynamics a system qudit interacts with an environment consisting of  $N$ qudits. Such interaction is modeled by pairwise collisions between the system qudit an a single environment   qudit. Each collision is described by a random unitary operator. The  environment,  i.e. $N$, is assumed to be so large that the system never collides twice with the same environment qudit. In pictorial terms one can think of a single qudit colliding in sequence with the individual  qudits of a long chain. 
The overall state of the system and environment,  after $t$ collisions, is 
\begin{equation}
\varrho ^{(t)}_{SE} = U_{0t} \cdots U_{02}U_{01} \varrho^{(0)}_{SE} U^{\dagger}_{01} U^{\dagger}_{02}\cdots U^{\dagger}_{0t}, 
\label{eq:uni}
\end{equation}
where $U_{0j}$ is a random unitary operator acting on the pair of qudits $0,j$;  $0$ labels the system   qudit and $j=1,\cdots ,N$ labels the environment qudits. Let us assume that the system and environment are in an initial tensor product state $\varrho_{SE}^{(0)} = \varrho_{S}^{(0)} \eta_1  \eta_2 \cdots \eta_N$. Since the collision operators  are random unitaries, the specific states $\eta_i$ are irrelevant and we can assume, without loss of generality that all the environment qudits  are in the same initial state $\eta$. 

In order to  characterize the relaxation process we first 
consider the decay of the system purity after $t$ collisions. 
We remind the reader that, given a density operator  $\varrho$, its purity is defined as $\mathcal{P}=\hbox{Tr}\left[{\varrho^{2}}\right]$.
The purity is a decreasing function of the degree of statistical mixture
of $\varrho$ and, for a qudit of dimension $\mu$ takes values in the range
$ \frac{1}{\mu} \leq\mathcal{P}\leq 1$,
where $ \mathcal{P}=1$ corresponds to pure states
and $\mathcal{P}=\frac{1}{\mu}$ to the completely unpolarized mixed state.  
Since we are focussing our attention to the purity of the system, our model is equivalent to a system   qudit colliding with a single environment   qudit whose state is refreshed to its initial state $\eta$ after each collision. We will show that, after averaging over random unitary collisions, 
$\mathcal{P}(t)$ can be analytically calculated.

\subsection{Colliding qubits} 

Let us first consider the case in which both system and environment consist of qubits.
The density operator of the system and environment qubits  can be written as

\begin{equation}
\varrho_{SE}=\sum_{\alpha_0,\alpha_E}c_{\alpha_0\alpha_E}
\sigma_0^{\alpha_0}\otimes \sigma_E^{\alpha_E},
\label{eq:rhoSE}
\end{equation}
where $\sigma_0^{\alpha_0}$  ($\sigma_E^{\alpha_E}$) denotes a Pauli matrix
acting on the system (environment) qubit, with  $\alpha_i\in\{0,x,y,z\}$ and  $\sigma^0=I$.
The purity of the overall system and environment after $t$ collisions then reads
\begin{equation}
\mathcal{P}_{SE}=4\sum_{\alpha_0\alpha_1} c_{\alpha_0\alpha_E}^2(t)
\end{equation}
and the system's purity is given by
\begin{equation}
\mathcal{P}(t)=8\sum_{\alpha_0} c_{\alpha_0 0}^2(t).
\end{equation}
Note that $\mathcal{P}_{SE}$ is invariant under unitary evolution, i.e., the overall system-environment purity 
is the same before and after each collision (of course before the state of the environment qubit is refreshed).
The constraints 
\begin{equation}
\hbox{Tr}\left[\varrho_{S E}\right]=1,
\quad
\hbox{Tr}\left[\varrho_{S E}^2\right]=\mathcal{P}_{SE}
\label{constrains}
\end{equation} 
lead do
\begin{equation}
c_{00}=\frac{1}{4},
\quad
\sum_{(\alpha_0,\alpha_1)\ne(0,0)} c_{\alpha_0\alpha_1}^2 =
\frac{4 \mathcal{P}_{SE}-1}{16}.
\end{equation}

It has been shown~\cite{Znidaric2007} that when two qubit collide with a sequence of random $U(4)$ unitaries the ensemble averaged coefficients $c^2(t)$ evolve according to a Markov chain as
\begin{equation}
c^2(t+1) = c^2(t)M, 
\label{Mc}
\end{equation}
where  
\begin{equation}
c^2=(c_{00}^2,c_{0x}^2,...,c_{zz}^2)
\end{equation}
and
\begin{equation}
M=\left(
\begin{array}{cccc}
 1 & 0 &\cdots & 0 \\  
 0 & \frac{1}{15} & \cdots& \frac{1}{15} \\ 
 \vdots & \vdots  &\ddots & \vdots \\
 0 & \frac{1}{15} & \cdots & \frac{1}{15} 
\end{array}
\right)
\label{M}
\end{equation}
is a  Markov $16\times 16$ matrix. The structure of $M$ is clear: it leaves $c^2_{00}$ unchanged while all the other components of $c^2$ are uniformly mixed.
The equilibrium state of such chain, if the state of the environment qubit is {\em not} refreshed after each collision,  must lie in the subspace spanned by the eigenvectors of $M$ with eigenvalue $1$. 
A normalized basis corresponding to the unit eigenvalue of $M$ is given by 
\begin{equation}
v_0=(1,0,...,0),\quad
v_1=\frac{1}{\sqrt{15}}(0,1,...,1).
\end{equation}
The equilibrium state then reads 
\begin{equation}
v=x_0v_0+x_1v_1,
\end{equation}
where
\begin{eqnarray*}
x_0&=&\langle v_0, v \rangle = c_{00}^2=\frac{1}{16},
\\
x_1&=&\langle v_1, v \rangle = \frac{1}{\sqrt{15}}
\sum_{(\alpha_0,\alpha_1)\ne(0,0)}c_{\alpha_0\alpha_1}^2
=\frac{1}{\sqrt{15}}\frac{4\mathcal{P}_{SE}-1}{16}.
\end{eqnarray*}
Therefore,
\begin{equation}
v=\left(\frac{1}{16},\frac{4\mathcal{P}_{SE}-1}{240},...,
\frac{4\mathcal{P}_{0E}-1}{240}\right).
\label{vv}
\end{equation}
Finally, we obtain
\begin{equation}
\mathcal{P}=8\sum_{\alpha_0}c_{\alpha_0 0}^2=
\frac{1}{2}+\frac{1}{10}(4\mathcal{P}_{SE}-1).
\label{P}
\end{equation}
For pure system-environment states ($\mathcal{P}_{SE}=1$) we
recover Lubkin's results~\cite{Lubkin1978}, 
\begin{equation}
\mathcal{P}=\mathcal{P}_L=\frac{4}{5}.
\end{equation}
It is important to note that such (ensemble averaged) 
equilibrium state is reached after a single random collision. In other words, as it should be, the state of the system and of the first environment qubit after a single collision is a two-qubit random state.

We now consider the case in which the state of the environment qubit is refreshed after each collision. After $t$ collision Eq.~(\ref{P}) becomes 
\begin{equation}
\mathcal{P}(t)=
\frac{1}{2}+\frac{1}{10}[4\mathcal{P}_{SE}(t-1)-1],\;
(t=1, 2, ...) .
\end{equation}
Note however that  ${\cal P}_{SE}$  changes when the environment qubit is reset to its initial state $\eta$. 
Just before the $t$-th collision we have
\begin{equation}
\mathcal{P}_{SE}(t-1)=\mathcal{P}(t-1)\mathcal{P}_{\eta},
\end{equation}
where $\mathcal{P}_{\eta}$ is the purity of the unperturbed environment qubit state $\eta$.

We now solve the equation
\begin{equation}
\mathcal{P}(t)=\frac{1}{2}+\frac{1}{10}[4\mathcal{P}(t-1)\mathcal{P}_{\eta} -1],\;
(t=1,2,,...).
\end{equation}
The equilibrium value 
\begin{equation}
\mathcal{P}(\infty)=\frac{2}{5 - 2\mathcal{P}_{\eta}}
\end{equation} 
is obtained as the solution to the equation
\begin{equation}
\mathcal{P}(\infty)=\frac{1}{2}+\frac{1}{10}[4\mathcal{P}(\infty)\mathcal{P}_{\eta}-1].
\end{equation}
If we define $\xi(t)=\mathcal{P}(t)-\mathcal{P}(\infty)$,
we obtain
\begin{equation}
\xi(t)=\frac{2}{5}\mathcal{P}_{\eta}\xi(t-1)=\left(\frac{2}{5}\mathcal{P}_{\eta}\right)^t\xi(0).
\label{rate}
\end{equation}
Note that  $\mathcal{P}(\infty)$ as well as the decay rate to equilibrium depend on $\mathcal{P}_{\eta}$: the lowest the purity of the environment qubits the fastest is the approach to equilibrium and the lowest the equilibrium value of the system purity. Although this has some analogies with the relaxation to thermal equilibrium it is worth stressing that the two processes have some important differences. In particular note that, contrary to the case of thermalization, $\mathcal{P}(\infty) \neq \mathcal{P}_{\eta}$.  Furthermore also the dependence on $\mathcal{P}_{\eta}$ of the decay rate is not the one expected for the relaxation to equilibrium. This is not surprising as collisions modeled by random collisions are not elastic. Indeed it has been shown in~\cite{Scarani2002} that the only operator guaranteeing thermalization (and in general homogenization~\cite{Ziman2002}) is the partial swap.

\begin{figure}[htbp!]
\begin{center}
\includegraphics[width=7cm]{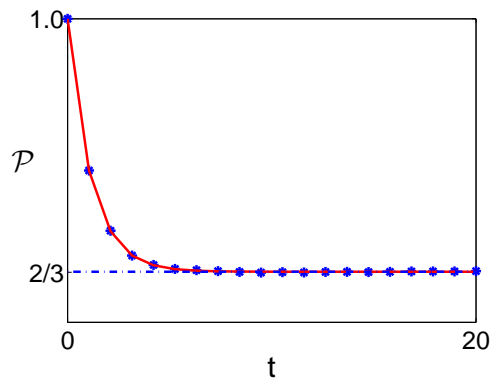}
\end{center}
\caption{(color online) Ensemble averaged purity functional behavior for $\mu=2,\nu=2$. The limit value is $\mathcal{P}(\infty)=2/3$. The markers are the ensemble averaged values. The curve shows the analytic result of 
Eq.~(\ref{eq:markovdecay22})}
\label{fig1}
\label{2per2}
\end{figure}

When the chain qubits are initially in a pure state, we have 
\begin{equation}
 \mathcal{P}(\infty) =\frac{2}{3}
 \end{equation}
 and equation (\ref{rate})  becomes
\begin{equation}
\mathcal{P}(t)=\left(\frac{2}{5}\right)^t\,
\left(\mathcal{P}(0)-\frac{2}{3}\right)+\frac{2}{3}=
e^{-\lambda t}
\left(\mathcal{P}(0)-\frac{2}{3}\right)+\frac{2}{3},
\label{eq:markovdecay22}
\end{equation}
where $\lambda=\ln(5/2)=0.916...$ is the rate of approach to equilibrium.
Starting from a pure state we obtain 
$\mathcal{P}(1)=4/5$,
$\mathcal{P}(2)=54/75=0.72$,
$\mathcal{P}(3)=258/375\approx 0.688$, ..., in 
agreement with our numerical data, as shown in Fig.~\ref{2per2}.

\subsection{Colliding qudits}

The above Markov chain technique can be generalized to analyze the more general case in which a system qudit of dimension $\mu$ collides with a large number of environment qudits of dimension $\nu$. In this case system-environment interactions are modeled by random unitaries 
drawn from the $U(L)$-invariant Haar measure in
the overall Hilbert space of size $L=\mu\nu$. The steps of the analysis 
previously carried on for qubits can be straightforwardly followed to obtain 
the equilibrium value and the decay rate of the purity of colliding qudits. 
We use again representation (\ref{eq:rhoSE}), where the $\sigma$'s are now generalized Pauli matrices~\cite{Schlienz1995,Gottesman1999}. 
The purity of the overall system then reads
\begin{equation}
\mathcal{P}_{SE}=
\mu \nu\sum_{\alpha_0=0}^{\mu-1}
\sum_{\alpha_1=0}^{\nu-1} c_{\alpha_0\alpha_1}^2(t)
\end{equation}
and the system's purity is given by
\begin{equation}
\mathcal{P}(t)=\mu \nu^2 \sum_{\alpha_0} c_{\alpha_0 0}^2(t).
\end{equation}
The constraints
\begin{equation}
\hbox{Tr}\left[\varrho_{S E}\right]=1,
\quad
\hbox{Tr}\left[\varrho_{S E}^2\right]=\mathcal{P}_{SE}
\end{equation}
lead do
\begin{equation}
c_{00}(t)=\frac{1}{\mu \nu},
\quad
\sum_{(\alpha_0,\alpha_1)\ne(0,0)} c_{\alpha_0\alpha_1}^2(t)=
\frac{\mu\nu \mathcal{P}_{SE}-1}{(\mu\nu)^2}.
\end{equation}
Also in this case the vector
\begin{equation}
c^2 = (c_{00}^2, \cdots, c_{\mu \nu}^2)
\end{equation}
evolves in time according to a Markov chain like in Eq.~(\ref{Mc}).
We extend Znidari\v c's conjecture~\cite{Znidaric2007} 
about the form of the Markov matrix
by assuming again that $M$ leaves $c^2_{00}$ unchanged while it mixes uniformly all other components. $M$ must therefore be a $(\mu\nu)\times(\mu\nu)$ matrix of the form

\begin{equation}
M=\left(
\begin{array}{cccc}
 1 & 0 &\cdots & 0 \\  
 0 & \frac{1}{(\mu\nu)^2-1} & \cdots& \frac{1}{(\mu\nu)^2-1} \\ 
 \vdots & \vdots  &\ddots & \vdots \\
 0 & \frac{1}{(\mu\nu)^2-1} & \cdots & \frac{1}{(\mu\nu)^2-1}
\end{array}
\right).
\label{Mmi}
\end{equation}

The two-dimenisonal 
eigenspace corresponding to the unit eigenvalue 
of $M$ is spanned by the vector basis 
\begin{equation}
v_0=(1,0,...,0),\quad
v_1=\frac{1}{\sqrt{(\mu\nu)^2-1}}(0,1,...,1).
\end{equation}
Following the same steps that lead to 
Eq.~(\ref{vv}) we obtain that the equilibrium state is 
\begin{equation}
v=\left(\frac{1}{(\mu\nu)^2},\frac{\mu\nu\mathcal{P}_{SE}-1}{(\mu\nu)^2
[(\mu\nu)^2-1]},...,
\frac{\mu\nu\mathcal{P}_{SE}-1}{(\mu\nu)^2[(\mu\nu)^2-1]}\right).
\end{equation}
The system purity of such state is 
\begin{equation}
\mathcal{P}=\mu\nu^2\sum_{\alpha_0}c_{\alpha_0 0}^2=
\frac{1}{\mu}+\frac{\mu^2-1}{\mu[(\mu\nu)^2-1]}(\mu\nu\mathcal{P}_{SE}-1).
\end{equation}
If the initial system-environment state is pure ($\mathcal{P}_{SE}=1$) we
recover Lubkin's result~\cite{Lubkin1978}: 
\begin{equation}
\mathcal{P}=\frac{\mu+\nu}{\mu\nu+1}.
\end {equation}

This means again that after a single collision the state of the  system and of the first colliding qudit is a random state in the Hilbert space of dimension $\mu\otimes\nu$

\begin{figure}[htbp]
\includegraphics[width=7cm] {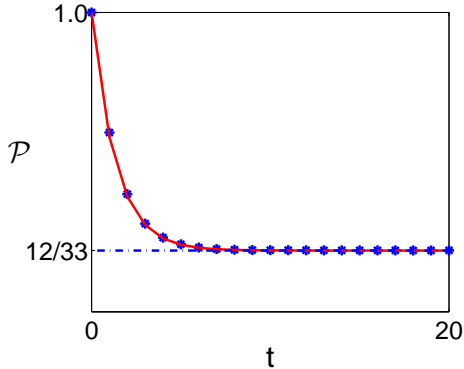}
\caption{(color online) Ensemble averaged purity functional behavior for $\mu=4,\nu=2$. The limit value is $\mathcal{P}(\infty)=12/33$. The markers are the ensemble averaged values. The curve shows the analytic result of 
Eq.~(\ref{eq:markovdecay42})}
 \label{4per2}
\end{figure} 

\begin{figure}[htbp!]
\begin{center}
\includegraphics[height=3.2cm]{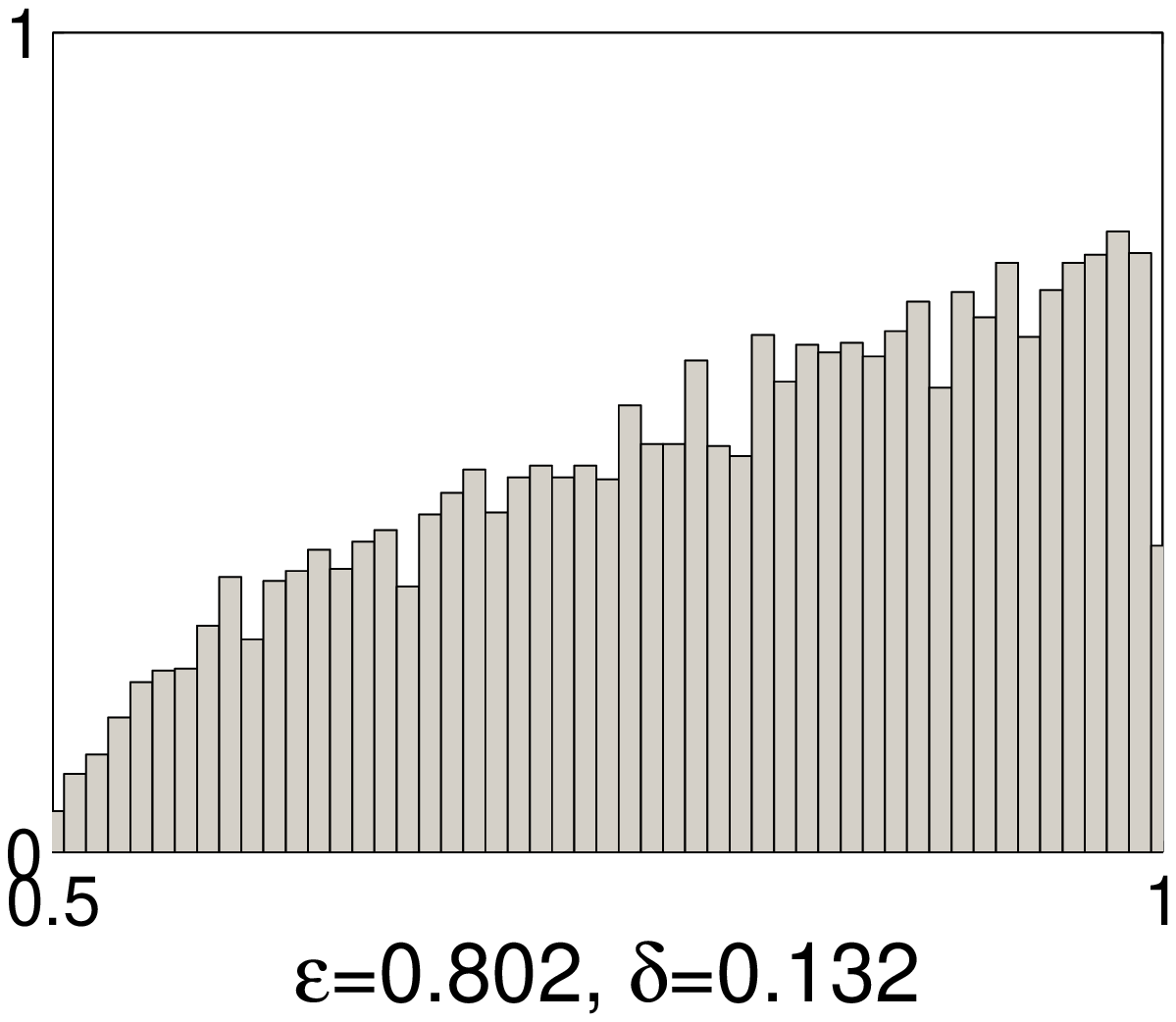}
\includegraphics[height=3.2cm]{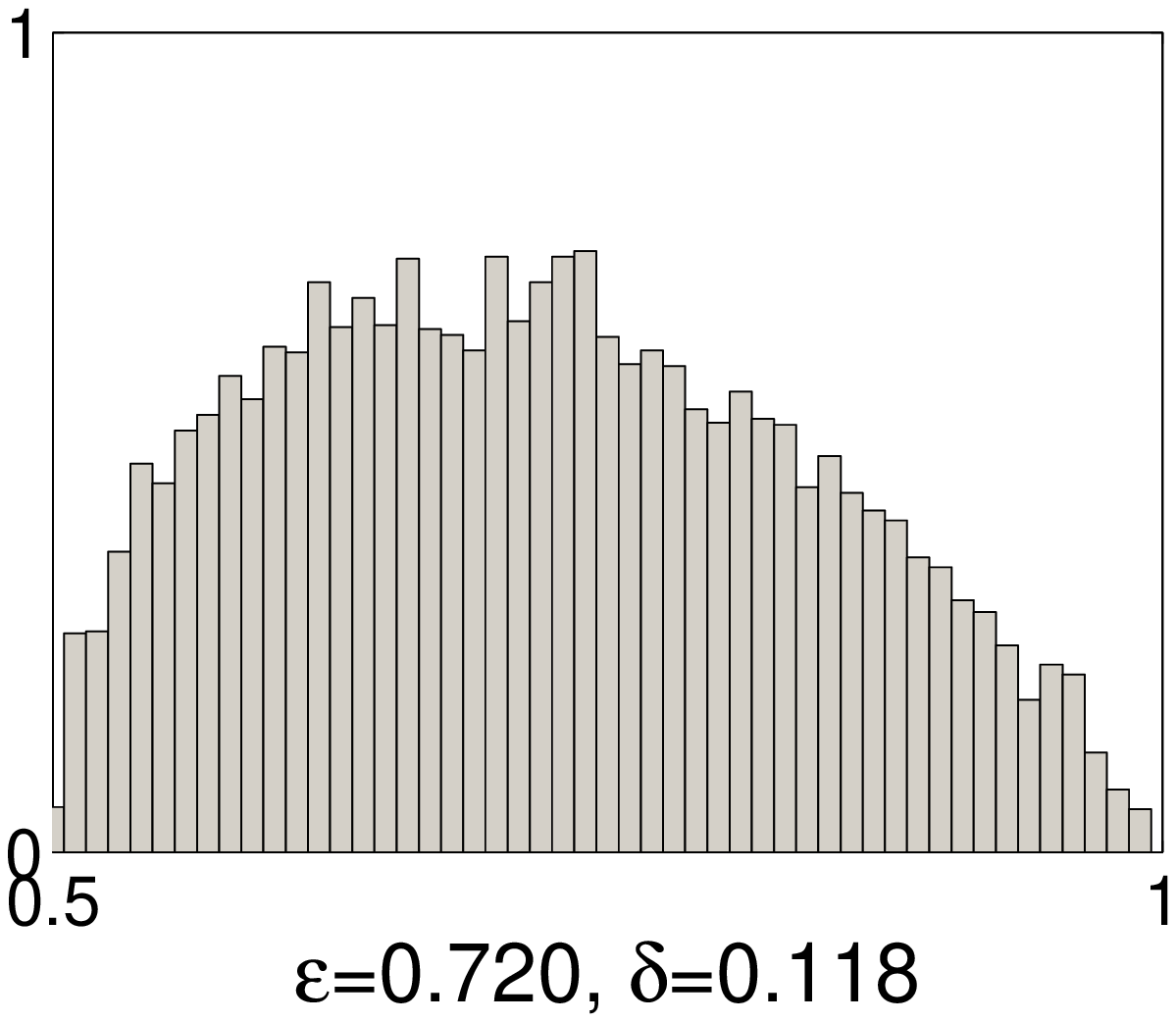}
\\
\includegraphics[height=3.2cm]{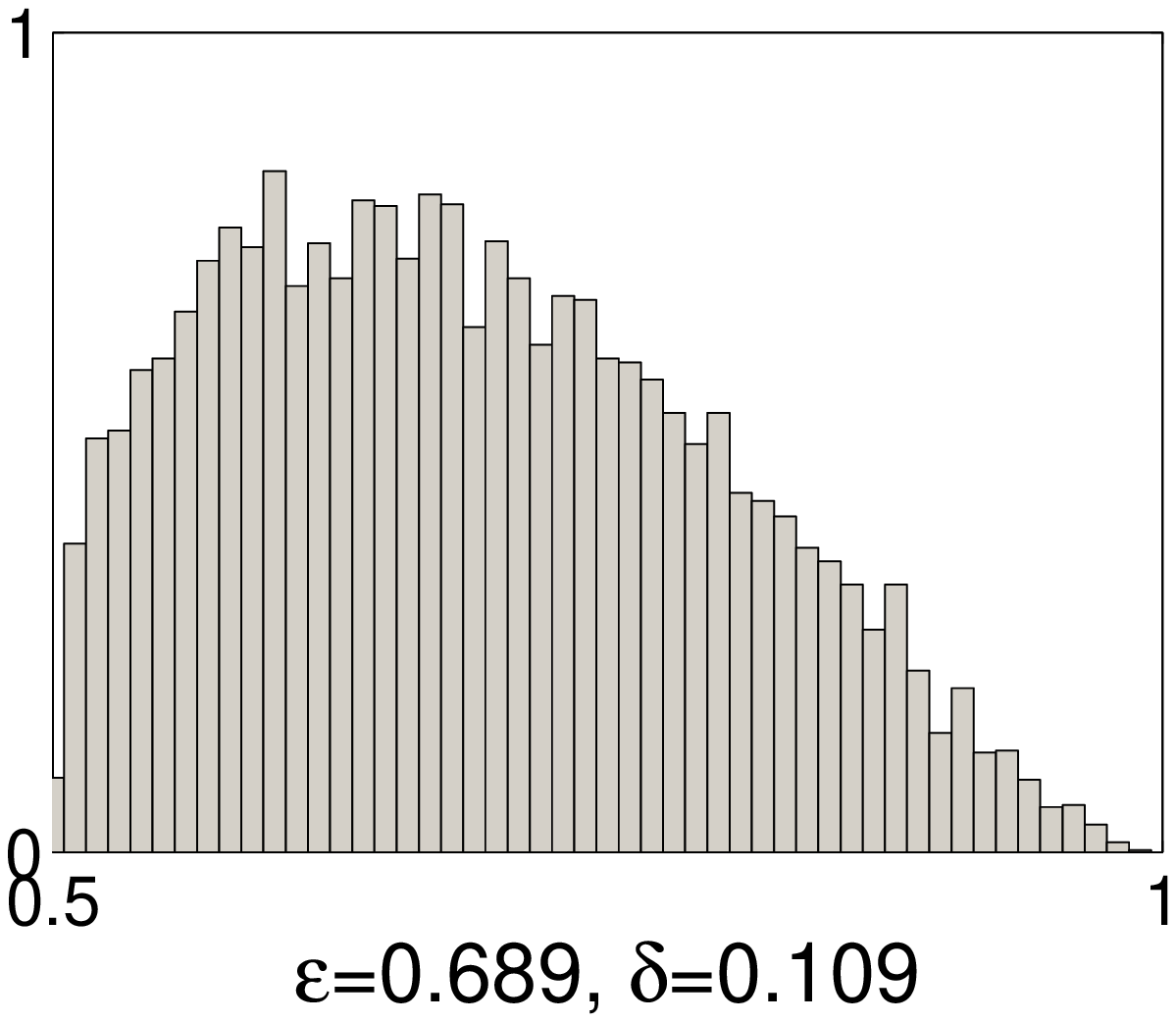}
\includegraphics[height=3.2cm]{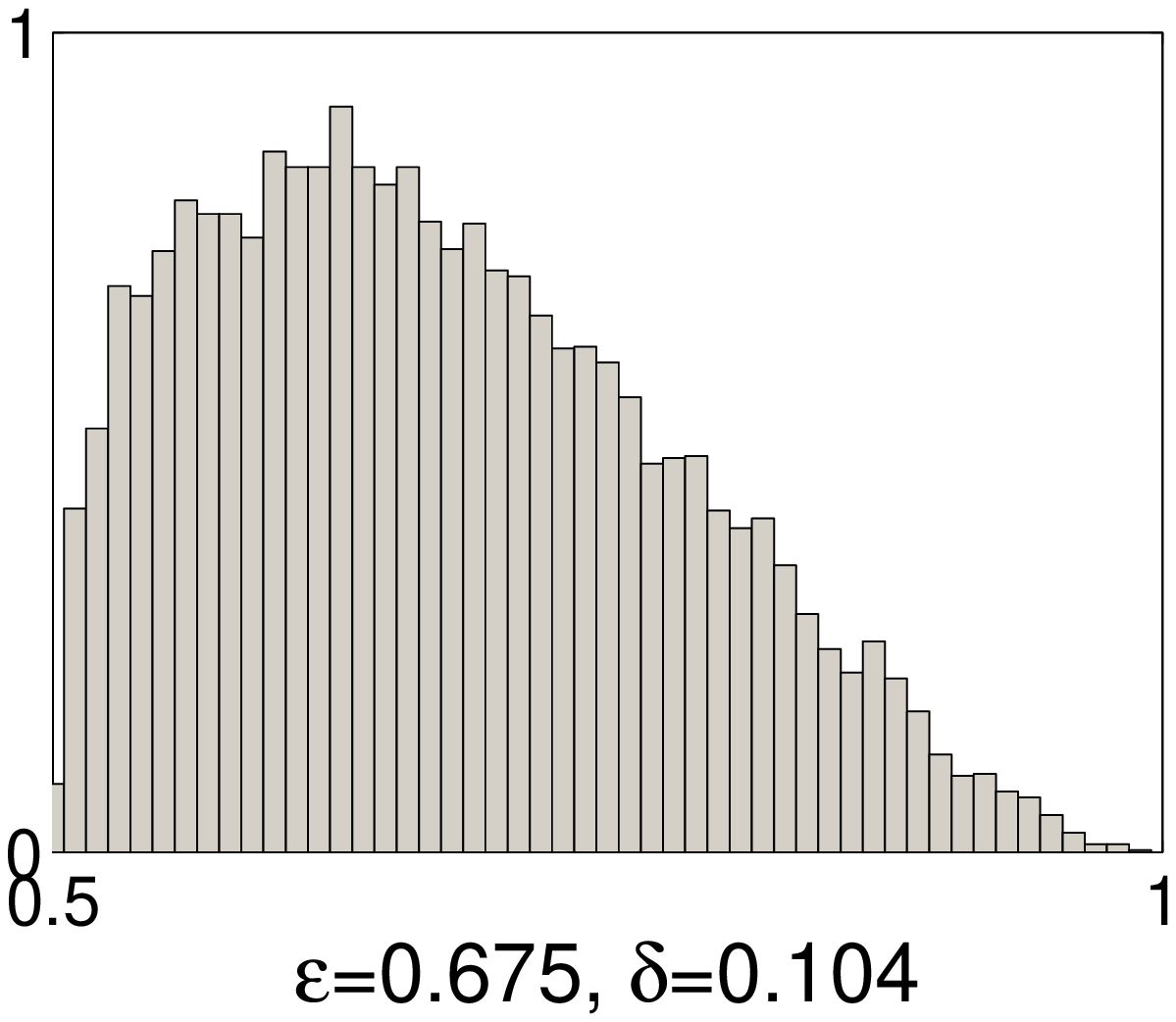}
\\
\includegraphics[height=3.2cm]{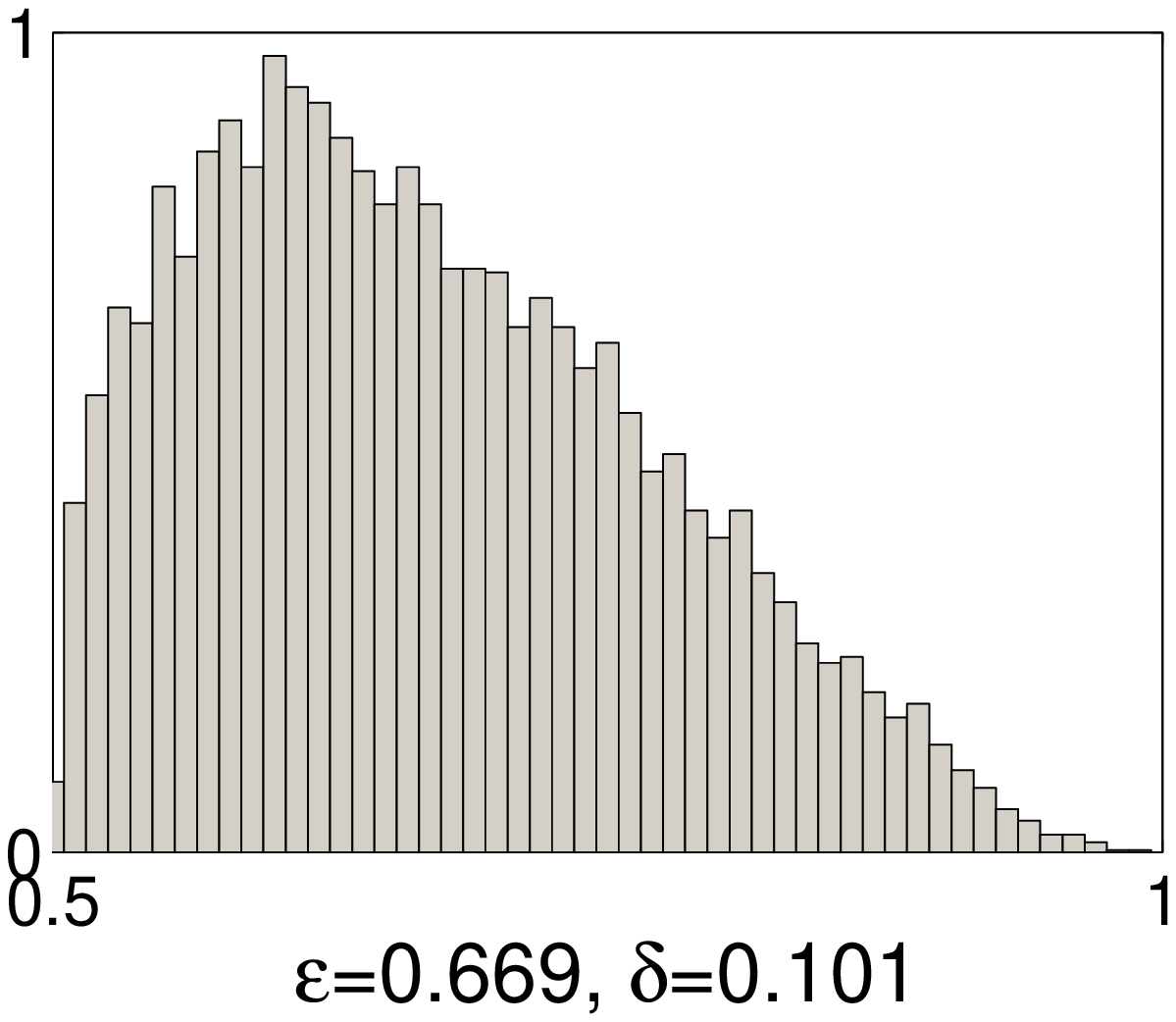}
\includegraphics[height=3.2cm]{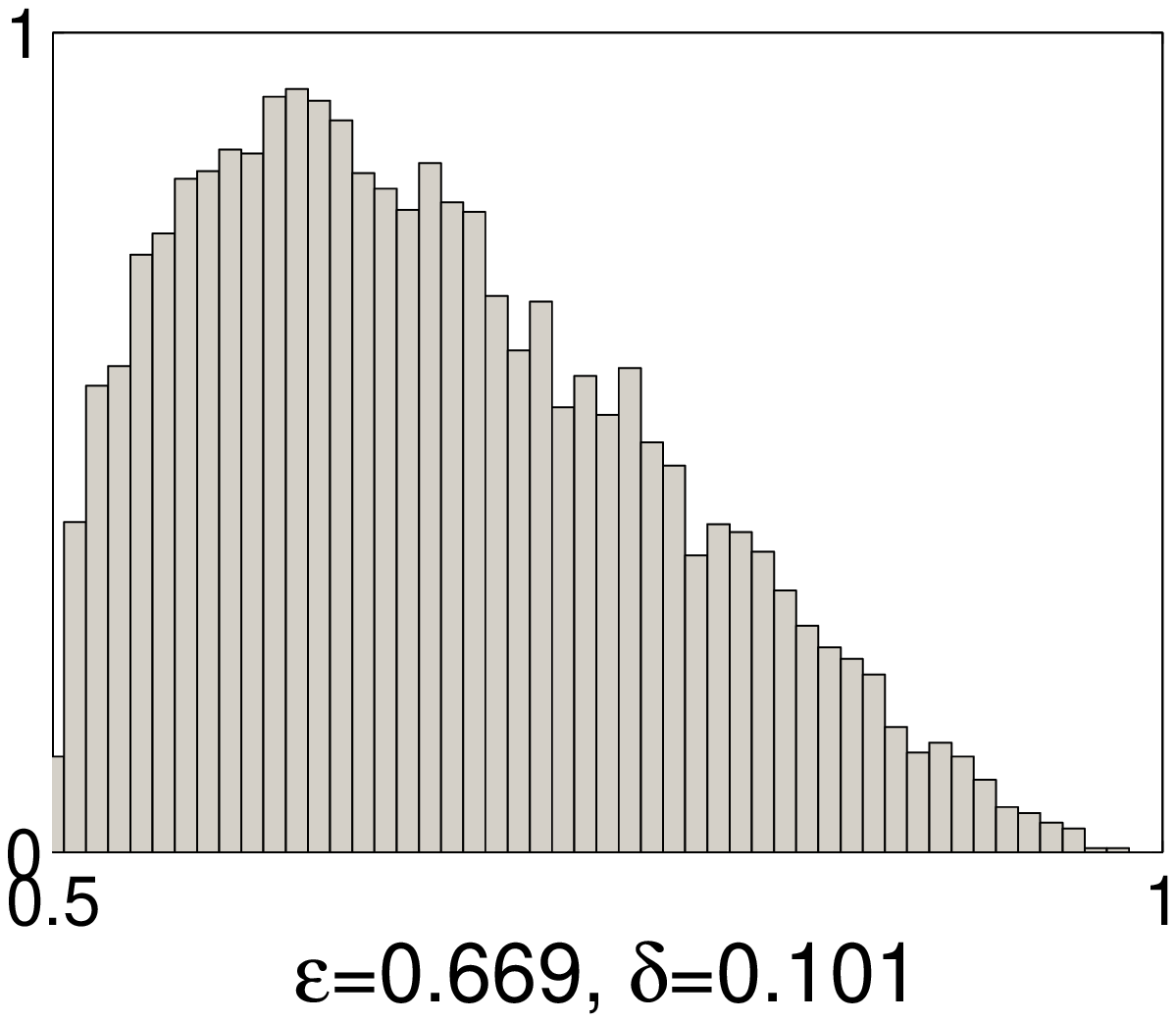}
\end{center}
\caption{From top to bottom, reading from left to right: $\mathcal{P}(1),\mathcal{P}(2),\mathcal{P}(3),\mathcal{P}(4),\mathcal{P}(5),\mathcal{P}(6)$ statistical distributions. 
In the lower side of each figure are shown the statistical average (indicated as $\epsilon$) and the standard deviation (indicated as $\delta$) values.}
\label{hist2per2}
\end{figure}

If the environment is refreshed after each collision, we have
\begin{equation}\label{pur}
\mathcal{P}(t+1)=\frac{1}{\mu}+\frac{\mu^2-1}{\mu[(\mu\nu)^2-1]}
[\mu\nu\mathcal{P}(t)-1].\;
(t=0,1,...).
\end{equation}
This equation has steady state
\begin{equation} 
\mathcal{P}(\infty)=\frac{\mu+(\mu\nu)}{\mu(\mu\nu)+1}.
\label{steady}
\end{equation}
If we define $\xi(t)=\mathcal{P}(t)-\mathcal{P}(\infty)$,
we obtain
\begin{equation}
\xi(t)=\alpha\xi(t)=\alpha^t\xi(0),
\end{equation}
where 
\begin{equation}
\alpha\equiv \frac{\nu(\mu^2-1)}{(\mu\nu)^2-1}.
\end{equation}
Therefore,
\begin{equation}
\mathcal{P}(t)=e^{-\lambda t}
\left(\mathcal{P}(0)-\mathcal{P}(\infty)\right)+\mathcal{P}(\infty),
\label{eq:markovdecay42}
\end{equation}
with $\lambda=-\ln \alpha$.
In  Fig.~\ref{4per2}, $\mathcal{P}(\infty)$ for  $\mu=4,\nu=2$ is plotted.

\subsection{Purity statistics}

\begin{figure}[htbp]
\includegraphics[width=7cm] {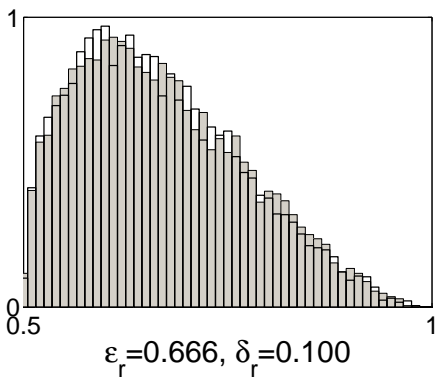}
\caption{Histograms comparison. Grey shadow: 
system purity distribution after the sixth collision (after which the equilibrium distribution is in 
practice already reached). Black: purity distribution of a system qubit colliding with a single qudit of dimension $\nu=4$, with average  $\epsilon_{r} = 0.666$ and variance $\delta_{r} = 0.1000$}
 \label{comp_ist}
\end{figure}

Eq.~(\ref{steady})  suggests  that the average system purity is the same one would obtain from a partition of a random state of a Hilbert space of dimension $\mu (\mu\nu)$, i.e. Eq.~(\ref{steady}) coincides with Lubkin's result if we assume that the system qubit has interacted with an environment of effective size $\nu_e = \mu\nu$. To support this conjecture we have plotted in Fig~\ref{hist2per2} the numerically generated histograms of the statistical distributions of $\mathcal{P}(t)$ for the first $6$ collisions for the case of colliding qubits.  
Indeed, as shown in Fig.~\ref{comp_ist}, after just six collisions 
the histogram practically coincides with the purity distribution 
of a system qubit colliding with a single qudit of dimension $4$. 

An intuitive explanation of such behavior can be given in simple terms. After the first collision between system and environment qudits the system becomes mixed - even if originally it was in a pure state. Following a standard mathematical procedure however its state can  be purified by introducing a fictitious qudit of dimension $\mu$ entangled with the system qudit. Therefore the whole process can be seen as a sequence of collisions among three qudits - the third one being the fictitious purification qudit - in an overall pure state. There is evidence~\cite{Gennaro2008} that in this case the overall system will evolve into a pure random sate in a Hilbert space of dimension $\mu^2\nu$. This explains the
purity statistics of the system qudit.

We have further numerical evidence to support the above  analysis: in~\cite{Scott2003} Scott and Caves reported an analytical expression of the average variance of the system purity distribution for a random interaction between a system of $\mu$ degrees of freedom and an environment of $\nu$ degrees of freedom. Such variance turns out to be 
\begin{equation} 
\langle \mathcal{P}^2 \rangle - \langle \mathcal{P} \rangle^2 
=\frac{2(\mu^{2}-1)(\nu^{2}-1)}{(\mu\nu+3)(\mu\nu+2)(\mu\nu+1)^{2}}.
\label{Scott}
\end{equation}
In our case instead the variance of the system purity distribution turns out to be well described by the above formula if we make the substitution 
$\nu\rightarrow  \nu_e=\nu\mu$ i.e.
\begin{equation} 
\langle \mathcal{P}^2 \rangle - \langle \mathcal{P} \rangle^2 
=\frac{2(\mu^{2}-1)(\nu_e^{2}-1)}{(\mu\nu_e+3)(\mu\nu_e+2)(\mu\nu_e+1)^{2}},
\label{Scott1}
\end{equation}
In the following Table~\ref{tavola} we report the standard deviations for different values of $\mu$ and $\nu$. In the second column we report the value of  the Scott \& Caves modified standard deviations of Eq.~(~\ref{Scott1}) while in the third column we report the corresponding numerically computed standard deviation values. 
Indeed a very good agreement between the two set of values is clearly seen. 
\begin{table}[htbp] 
\begin{tabular}{c|c|c}
$\mu, \nu$ & $\;$Scott \& Caves$\;$ &  $\;$Collision model$\;$\\
 & & \\
$\mu=2$, $\nu=2\;$ & $0.1005$ & $0.1010$\\
$\mu=2$, $\nu=3\;$ & $0.0769$ & $0.0767$\\
$\mu=3$, $\nu=2\;$ & $0.0608$ & $0.0628$\\
$\mu=4$, $\nu=2\;$ & $0.0382$ & $0.0388$\\
$\mu=2$, $\nu=4\;$ & $0.0618$ & $0.0632$\\
$\mu=3$, $\nu=3\;$ & $0.0433$ & $0.0438$\\
\end{tabular}
\caption{Comparison between the Scott \& Caves modified standard deviations $(\mu, \nu\rightarrow \mu, \nu_e=\nu\mu)$ and numerically calculated standard deviations}
\label{tavola}
\end{table}

\section{entanglement dynamics}

To further characterize the approach to equilibrium, in this section 
we illustrate some aspects of the entanglement dynamics
for qubits, i.e. $\mu=\nu=2$. Since the overall state of the system and
chain remains pure, the entanglement dyanamics is conveniently
characterized in terms of the so-called 
tangles~\cite{Wootters1998,Wootters2000}.
We remind the reader that,
given the density operator $\rho_{ij}$ of a
bipartite system of two qubits, the tangle $\tau_{i|j}$ is defined as
\begin{equation}
\tau_{i|j}(\rho)=[\max\left\{0,\alpha_{1}-\alpha_{2}-{\alpha_{3}}-{\alpha_{4}}\right\}]^2,
\end{equation}
where $\left\{\alpha_{k}\right\}$ ($k=1,..,4$) are the square roots of the eigenvalues (in non-increasing order) of the
non-Hermitian operator $\bar{\rho}_{ij}=
\rho_{ij}(\sigma_{y}\otimes\sigma_{y})
\rho_{ij}^{*}(\sigma_{y}\otimes\sigma_{y})$, $\sigma_{y}$ is the
$y$-Pauli operator and $\rho_{ij}^{*}$ is the complex conjugate
of $\rho_{ij}$, in the eigenbasis of the $\sigma_z  \otimes\sigma_z$ operator.
The concurrence $C$ is defined simply as 
$C_{ij}=\sqrt{\tau_{i|j}}$.
The tangle $\tau_{i|j}$, or equivalently the concurrence
$C_{ij}$ can be used to quantify the entanglement between the pair of
qubits $i,j$ for an arbitrary reduced density operator $\rho_{ij}$.  
In our case the overall
state of the system and the chain is pure. 
Therefore, the amount of entanglement
between qubit $i$ and all the remaining qubits can be quantified by the
tangle $\tau_{i| \mbox{rest}}=4\det\rho_i$.
After $t$ collisions, the tangle 
$\tau_{0|\mbox{chain}}(t)$ between the system qubit and the chain
conveys the same information as the purity $\mathcal{P}(t)$. 
Indeed, it is easy to show that
\begin{equation}
\tau_{0|\mbox{chain}}(t)=2-2\mathcal{P}(t).
\label{eq:tauP}
\end{equation}
The purely multipartite entanglement $\tau_M$ established between the system qubit
and the qubits of the chain can be quantified as  
\begin{equation}
\tau_M(t)=\tau_{0|\mbox{chain}}(t)
-\sum_{j=0}^t\tau_{0|j}.
\label{eq:tauref}
\end{equation}
i.e. as the amount of entanglement which cannot be ascribed to purely bipartite entanglement between the system qubit and each individual environment qubit. 
Due to the complexity of the analytical expressions of the tangles,
we resort to numerical simulations. Note that, in order to evaluate the tangle 
$\tau_{0|\mbox{chain}}$ we have to numerically find the pairwise 
tangle between the system qubit and each qubit of the chain. This implies that we have to 
retain the overall system-chain
density matrix; i.e. in contrast to the computation of the system purity
$\mathcal{P}$, we cannot trace over the chain after each collision.
In other words, as far as the entanglement dynamics is concerned, the 
qubit-chain model is not equivalent to a model in which the system
qubit collides with a single environment qubit whose state is refreshed
after each collision.

\begin{figure}[htbp]
\includegraphics[width=7cm] {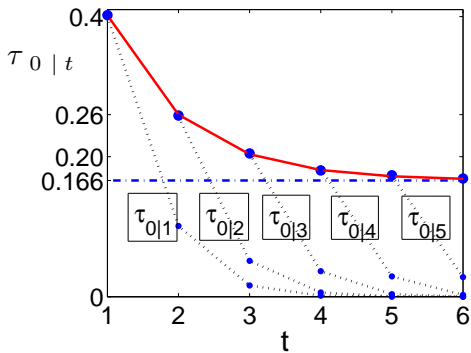}
\caption{(color online) Pairwise tangles $\tau_{0|j}$ as a function 
of the number $t$ of collisions (dashed curves with circles). 
The top circles refer to $\tau_{0|t}(t)$ immediately after 
the $t$-th collision and are fitted by the solid curve 
$\tau_{0|t}(t)-\tau_{0|\infty}(\infty)=0.512 \exp(-0.921 t)$,
with the fitting parameter $\tau_{0|\infty}(\infty)=0.166$.}
\label{tangle1}
\end{figure}

The pairwise tangles $\tau_{0|j}$ are shown in Fig.~\ref{tangle1},
as a function of the number of collisions. 
A non-zero tangle $\tau_{0|t'}$ between the system qubit ant the
$t'$-th qubit of the chain is generated immediately after the 
$t'$-th collision and then quickly decays for $t>t'$.
The dependence of $\tau_{0|t}(t)$ on the number $t$ of collisions 
if fitted satisfactorily by the exponential curve
$\tau_{0|t}(t)=0.166+0.512 \exp(-0.921 t)$,
suggesting the asymptotic value 
$\tau_{0|\infty}(\infty)=0.166$ for the paiwise tangle 
generated between the system qubit and a qubit of the chain
immediately after their collision.

\begin{figure}[htbp]
\includegraphics[width=7cm]{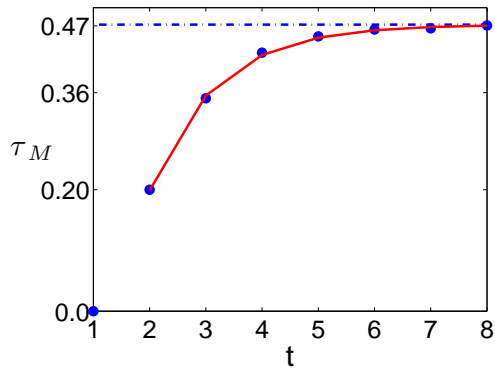}
\caption{(color online) Multipartite entanglement 
$\tau_M$ for the system qubit as a
function of the number $t$ of collisions. 
The solid line represents the
exponential fit 
$\tau_M(t)-\tau_M(\infty)
=-1.491 \exp(-0.849 t)$, with 
the fitting parameter $\tau_M(\infty)=0.472$.}
\label{fig:CKW}
\end{figure}

Our numerical data shown in Fig.~\ref{tangle1} also suggest that the
weight of the terms $\tau_{0|j}(t)$ with $j\ne t$ can be neglected
with respect to $\tau_{0|t}(t)$. 
Under this approximation and using 
Eqs.~(\ref{eq:tauref}) and (\ref{eq:tauP}) we can estimate the 
asymptotic multipartite entanglement established between the system qubit
and the qubits of the chain as 
\begin{equation}
\tau_M(\infty)
\approx[2-2{\mathcal P}(\infty)]
-\tau_{0|\infty}(\infty)\approx \frac{1}{2}.
\end{equation}
This expectation is confirmed by the numerical data shown in 
Fig.~\ref{fig:CKW}: the convergence of $\tau_M(t)$ 
to its asymptotic value $\tau_M(\infty)$ is well fitted
by the exponential decay
$\tau_M(t)-\tau_M(\infty)
=-1.491 \exp(-0.849 t)$, with 
the fitting parameter $\tau_M(\infty)=0.472
\approx \frac{1}{2}$.

\section{Conclusions}

We have investigates the dynamics of a system qudit of dimension $\mu$ sequentially interacting with the qudits, of dimension $\nu$, of a chain. Each pairwise collision is modeled as a random transformation drawn from the Haar measure on $U(\mu\nu)$. 
The relaxation to equilibrium, in terms of ensemble average over random  collisions, in analytically investigated by means of a Markov chain approach. We have shown that the steady state is the one corresponding to                
the steady state for random collisions with a single environment
qudit of effective dimension $\nu_e=\nu\mu$.
Furthermore, in contrast to the case of
the homogeneization process induced by purely
elastic partial swap collisions~\cite{Ziman}, random unitary collisions
can generate multipartite entanglement.

\section*{Acknowledgments}
 
G.G. and G.M.P. acknowledge support from   PRIN 2006 "Quantum noise in mesoscopic systems"  
and from EUROTECH S.p.A.


\begin{thebibliography}{0}
   
\bibitem{Scarani2002}
V.~Scarani, M.~Ziman, P.~\v{S}telmachovi\v{c}, N.~Gisin, and V.~Bu\v{z}ek, Phys. Rev.  Lett. {\bf 88}, 097905 (2002).

\bibitem{Ziman2002}
M.~Ziman, P.~\v{S}telmachovi\v{c}, V.~Bu\v{z}ek, M.~Hillery, V.~Scarani, and N.~Gisin, Phys. Rev. A{\bf 65}, 042105, (2002).
 
\bibitem{Ziman}
M.~Ziman, P.~\v{S}telmachovi\v{c}, and V.~Bu\v{z}ek,
J. Opt. B: Quantum Semiclassical Opt. {\bf 5}, S439 (2003);
M.~Ziman, P.~\v{S}telmachovi\v{c}, and V.~Bu\v{z}ek, Open Sys. \& Information Dyn. {\bf 12}, 81 (2005);
M.~Ziman and  V.~Bu\v{z}ek, Phys. Rev. A {\bf 72}, 022110, (2005).

\bibitem{Koniorczyk}
M.~Koniorczyk, A.~Varga, P.~Rap\v{c}an, and V.~Bu\v{z}ek,
Phys.Rev A {\bf 77}, 052106 (2008). 
 
\bibitem{Benenti2007}
G.~Benenti G. and G.M.~Palma,
Phys. Rev. A {\bf 75}, 052110 (2007).
 
\bibitem{Gennaro2008}
G.~Gennaro, G.~Benenti, and G.M.~Palma,
EPL {\bf 82}, 20006 (2008).  

\bibitem{Diaconis2005}
P.~Diaconis,
Notices of the AMS {\bf 52}, 11 (2005).  
 
\bibitem{Mezzadri2007}
F.~Mezzadri,
Notices of the AMS {\bf 54}, 592 (2007).

\bibitem{Kus1991}
M.~Kus and K.~Zyczkowski,
Phys. Rev. A {\bf 44}, 956 (1991).

\bibitem{Kus1994}
K.~Zyczkowski and M.~Kus,
J. Phys. A: Math. Gen. {\bf 27}, 4235 (1994).

\bibitem{Kus1996}
K.~Zyczkowski K. and M.~Kus,
Phys. Rev. E {\bf 53}, 319 (1996).
  
\bibitem{Kus1998}
M.~Pozniak, K.~Zyczkowski, and M.~Kus,
 J. Phys. A: Math. Gen. {\bf 31}, 1059 (1998).

\bibitem{Pineda2007}
C.~Pineda and T.H.~Seligman , 
Phys.Rev.A {\bf75}, 012106, (2007); 
C.~Pineda, T.~Gorin, and T.H.~Seligman,
New J. Phys. {\bf 9}, 106 (2007).

\bibitem{Petruccione2007}
A.~Akhalwaya, M.~Fannes, and F.~Petruccione,
J. Phys. A: Math. Theor. {\bf 40}, 8069 (2007). 
 
\bibitem{Emerson2003}
J.~Emerson, Y.S.~Weinstein, M.~Saraceno, S.~Lloyd, and D.G.~Cory,
Science {\bf 302}, 2098 (2003).
 
\bibitem{Emerson2004}
J.~Emerson,
AIP Conf. Proc. {\bf 734}, 139 (2004).
 
\bibitem{Emerson2005}
J.~Emerson, E.~Livine, and S.~Lloyd,
Phys. Rev. A {\bf 72}, 060302(R) (2005).

\bibitem{Weinstein2005}
Y.S.~Weinstein and C.S.~Hellberg,
Phys. Rev. Lett. {\bf 95}, 030501 (2005).
 
\bibitem{Plenio2007}
O.C.O.~Dahlsten, R.~Oliveira, and M.B.~Plenio, 
J. Phys. A: Math. Theor.{\bf 40}, 8081 (2007).

\bibitem{Oliveira2007}
R.~Oliveira, O.C.O.~Dahlsten, and M.B.~Plenio,
Phys. Rev. Lett. {\bf 98}, 130502 (2007).

\bibitem{Znidaric2007}
M.~\v Znidari\v c,
Phys. Rev. A {\bf 76}, 012318 (2007).
 
\bibitem{Rossini2008}
D.~Rossini and G.~Benenti,
Phys. Rev. Lett. {\bf 100}, 060501 (2008). 

\bibitem{Benenti2008}
G.~Benenti,
preprint arXiv:0807.4364v1 [quant-ph].

\bibitem{Znidaric2008}
M.~\v Znidari\v c,
Phys. Rev. A {\bf 78}, 032324 (2008).
 
\bibitem{Harrow2004}
A.~Harrow, P.~Hayden, and D.~Leung,
Phys. Rev. Lett. {\bf 92}, 187901 (2004).
 
\bibitem{Bennett2004}
C.H.~Bennett, P.~Hayden, D.~Leung, P.~Shor, and A.~Winter,
IEEE Trans. Inf. Theory {\bf 51}, 56 (2005).

\bibitem{Hayden2004}
P.~Hayden, D.~Leung, P.~Shor, and A.~Winter,
Commun. Math. Phys. {\bf 250}, 371 (2004).

\bibitem{Hayden2006}
P.~Hayden, D.W.~Leung, and A.Winter,
Commun. Math. Phys. {\bf 265}, 95 (2006).
 
\bibitem{Lubkin1978}
E.~Lubkin,
J. Math. Phys.{\bf 19}, 1028 (1978).

\bibitem{Schlienz1995}
J.~Schlienz and G.~Mahler,
Phys. Rev. A {\bf 52}, 4396 (1995).

\bibitem{Gottesman1999}
D.~Gottesman,
Chaos, Solitons \& Fractals {\bf 10}, 1749 (1999).

\bibitem{Wootters1998}
W.K.~Wootters,
Phys. Rev. Lett. {\bf 80}, 2245 (1998). 
 
\bibitem{Wootters2000}
V.~Coffman, J.~Kundu, and W.K.~Wootters,
Phys. Rev. A {\bf 61} 052306 (2000).

\bibitem{Scott2003}
A. Scott and C. Caves,
J. Phys. A: Math. Gen. {\bf 36} 9553 (2003).
 
\end{thebibliography}
\end{document}